%% file: sample-sigconf-authordraft.tex
\documentclass[sigconf]{acmart}
\usepackage{hyperref}
\usepackage{booktabs} 
\usepackage{caption}
\usepackage{graphicx, subfig}

\setcopyright{rightsretained}

\copyrightyear{2018} 
\acmYear{2018} 
\setcopyright{acmcopyright}
\acmConference[BDET 2018]{2018 International Conference on Big Data Engineering and Technology}{August 25--27, 2018}{Chengdu, China}
\acmBooktitle{2018 International Conference on Big Data Engineering and Technology (BDET 2018), August 25--27, 2018, Chengdu, China}
\acmPrice{15.00}
\acmDOI{10.1145/3297730.3297743}
\acmISBN{978-1-4503-6582-6/18/08}

\begin{document}
\title{Big Data Platform Architecture Under The Background of Financial Technology}

\author{Yi Liu}
\authornote{Corresponding author.}
\affiliation{%
	\institution{School of Data Science and Technology, Heilongjiang University}
	\streetaddress{No.74 Xuefu Road}
	\city{Harbin}
	\state{Heilongjiang}
	\postcode{150080}
}
\email{20165113@s.hlju.edu.cn\\N1800620A@e.ntu.edu.sg}

\author{Jiawen Peng}
\affiliation{%
  \institution{College of Science, Jiangxi University of Science and Technology}
  \streetaddress{86 Hongqi Avenue}
  \city{Ganzhou}
  \state{Jiangxi}
}
\email{jiawenp@163.com}

\author{Zhihao Yu}
\affiliation{%
  \institution{School of Information and Software Engineering, University of Electronic Science and Technology,  City,China}
  \city{Chengdu}
  \country{China}}
\email{zhihao@163.com}

\renewcommand{\shortauthors}{B. Trovato et al.}

\begin{abstract}
With the rise of the concept of financial technology, financial and technology gradually in-depth integration, scientific and technological means to become financial product innovation, improve financial efficiency and reduce financial transaction costs an important driving force. In this context, the new technology platform is from the business philosophy, business model, technical means, sales, internal management and other dimensions to re-shape the financial industry. In this paper, the existing big data platform architecture technology innovation, adding space-time data elements, combined with the insurance industry for practical analysis, put forward a meaningful product circle and customer circle.
\end{abstract}

%
%



\maketitle

\input{samplebody-conf}

\end{document}

%% file: samplebody-conf.tex
\section{Introduction}

With the popularity of the Internet and the rapid development of emerging technologies, the concept of financial science and technology has emerged as the new outlet for the development of the financial industry after the Internet finance, providing an infinite imagination for financial product innovation and service upgrading \cite{ref-1}. As one of the important industries in the financial industry, the insurance industry, driven by the wave of financial science and technology, is in deep integration with emerging technologies such as big data.

The foundation of insurance is the law of large numbers, therefore, the insurance industry and the data can be said to be closely related \cite{ref-2}. But now the insurance industry has problems such as less sample data, less real-time data, incomplete internal data, and lower data quality and dimensions. The insurance industry how to effectively use the data and mining new data, related to the survival and development of the insurance industry. For now, most of the traditional data platforms in China have single function and single service, which are not combined with other mature technologies. However, there are many foreign successful examples of combining GIS technology with big data technology. For example, a U.S. headmaster uses space-time big data technology to improve the probability of finding out students who dropped out earlier and make early intervention.

Big data platform with big data vision and thinking, combined with the very mature GIS technology, focusing on insurance business model, business philosophy, product design, management processes, the financial science and technology to achieve the full upgrade and transformation of the insurance industry.

Spatio-temporal big data platform mainly contains spatio-temporal data information database, data visualization platform and big data mining technology. This platform is based on the new model of "platform + big data + integrated services", which integrates spatial basic data, public information data, data visualization, big data mining technology and big data visualization deduction technology to provide data processing for the insurance industry With forecasting, data model building, data application crowd function and application support services, it is easy to form a product-centric product circle and a client-centric customer circle.\cite{ref-3}

This paper has two major contributions: \\(1) a brief review of the application of big data technology in the field of financial technology;\\ (2) designed a platform for the application of big data in the insurance industry, and introduced the relevant mainstream technology. For example, Hadoop, HDFS, Spark, and Hive. 

\section{The role of the big data platform}
The first is credit risk assessment. In the traditional method, the bank's default risk assessment for corporate customers is mostly based on static data such as past credit data and transaction data. The biggest drawback of this approach is the lack of forward-looking. Because the important factors affecting corporate default are not only the credit situation of the company's history, but also the overall development of the industry and real-time business conditions. The involvement of big data means makes credit risk assessment closer to reality.

Integration of internal and external data resources is a prerequisite for big data credit risk assessment. In general, in the process of identifying customer needs, estimating customer value, judging customer pros and cons, and predicting customer default, commercial banks need to rely on customer-related information already in the bank, as well as pedestrian information collected by external agencies. Information, customer public evaluation information, business operation information, income and expenditure consumption information, social related information, etc.

The second is supply chain finance. Using big data technology, banks can form a relationship map between enterprises based on investment, holding, lending, guarantees, and the relationship between shareholders and legal persons, which is beneficial to affiliate analysis and risk control. Knowledge maps organically organize fragmented data by establishing links between data, making data more easily understood and processed by humans and machines, and facilitating search, mining, analysis, and the like.

In terms of risk control, the bank takes the core enterprise as the entry point and treats several key enterprises in the supply chain as a whole. Using the circle analysis model, we continuously observe the changes of communication communication data between enterprises, and compare the abnormal data with the baseline data to assess the health of the supply chain and provide reference for the enterprise's post-loan risk control.
\section{The Collision Between Insurance and Big Data }

With the widespread popularity and development of big data technology, financial big data applications have become a hot trend in the industry, in transaction fraud identification, precision marketing, black production prevention, consumer credit, credit risk assessment, supply chain finance, stock market forecast, stock price Forecasting, smart investment, fraud insurance identification, risk pricing and other specific businesses involving banks, securities, insurance and other fields have been widely used. The application analysis capability of big data is becoming the core competitive factor for the future development of financial institutions.

Undoubtedly, financial big data has broad prospects for development. However, financial big data applications are also faced with a series of constraints such as insufficient data asset management, difficulty in technological transformation, lack of industry standards, pressure on security control, and imperfect policy support. In order to promote the development and application of financial big data, we must start from the aspects of policy support guarantee, data management capability improvement, industry standardization construction and application cooperation innovation, continuously strengthen the application of basic capabilities, and continuously improve the industrial ecological environment.

\begin{figure}
	\includegraphics{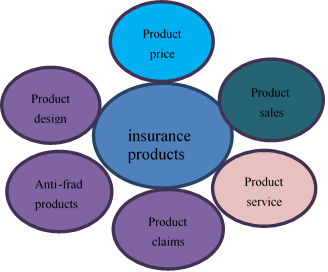}
	\caption{Sketch map of product circle.}
	\label{fig-1}
\end{figure}

The value of big data for the insurance industry goes without saying.
In the field of business innovation, how to find potential customers and find out the intrinsic requirements of customers \cite{ref-4}, to provide the most suitable service to customers is the key to the existence and development of the insurance industry. Space-time big data platform for the insurance industry to provide data and technical support to help the insurance industry to create product-centric products and customer geographies as the center of the customer circle, and promote the development of the industry, the formation of a "two centers a basic point" of development mode. (See Figure \ref{fig-1})

On operational costs, we collected data related to insurance business through big data technology and established a standardized and systematic insurance data system. On the one hand, the statistics of insurance customers' positioning and pricing of insurance products are easy to be eliminated, greatly reducing data collection The time and cost \cite{ref-5}. On the other hand, the use of relevant data to establish insurance risk aversion (especially in the field of health) platform to improve the ability of enterprises to control risk.

In precision pricing, big data will change the pricing rules of traditional insurance products. In the traditional insurance industry, actuaries based on the principle of large numbers, according to the principle of stochastic, extract some data from long-term and large-scale management practices, construct mathematical models and calculate insurance Rate, and further make product pricing \cite{ref-6}. In the space-time information base of big data platform, we dig out the relationship behind the data, through the algorithm in machine learning, can achieve accurate pricing, so as to maximize the benefits.

\begin{figure}
	\includegraphics{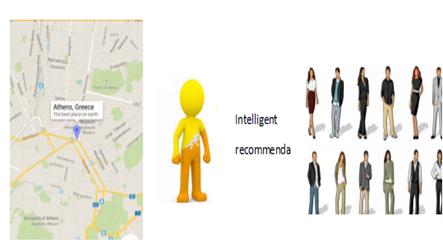}
	\caption{A map of customer circle based on location.}
	\label{fig-2}
\end{figure}
\section{The collision between insurance and big data}

Integrating the space-time information images of the taxpayers with the space-time remote sensing data, Internet data, business data and claims data of the insured in the overall business, based on which, the spatial-temporal data retrieval, space-time data analysis, Product service tracking, insurance fraud and other services.

The concept of "Big Data +" will be extended through such aspects as product design, product pricing, product recommendation\cite{ref-14}, product tracking service, product claim settlement and anti-fraud so as to broaden the application scenario of big data technology in smart insurance work, integrate data resources, customer needs, Product Features, Product Smart Product Recommendations, Wisdom Anti-Fraud.

From the relationship between the time and spatial relationship analysis of data, provide a variety of visual analysis capabilities to support the visual relationship between these three layouts, large amounts of data into valuable customer information resources\cite{ref-15}. Dig deeper into data resources to identify potential customers and cheat customers. Data visualization software provides platform capabilities for multi-dimensional visual data analysis, heuristic analysis process and data visualization capabilities \cite{ref-6}.

\section{The collision between insurance and big data }

\subsection{Big data platform positioning}

The big data platform mainly collects and processes data and real-time feedback to deal with huge amounts of structured data, semi-structured data and unstructured data \cite{ref-7}. Big data platform will be combined with the currently very mature GIS technology, business-oriented, providing a variety of data support services.

\subsection{Big data platform architecture}

\begin{figure}
	\includegraphics{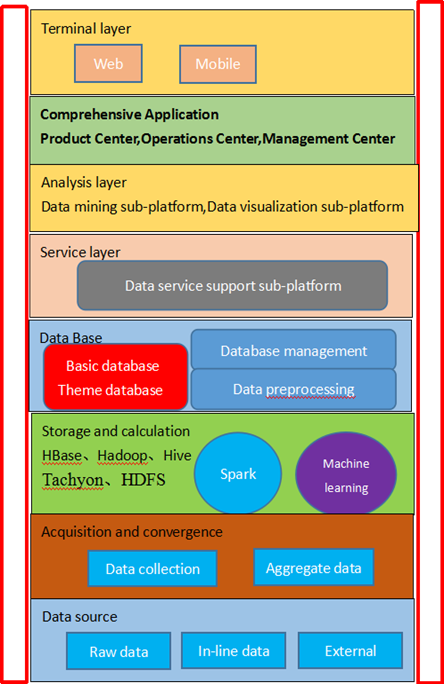}\label{fig-3}
	\caption{Big data platform overall framework.}
\end{figure}

The overall architecture of the platform is divided into seven layers, including data sources, data acquisition and aggregation, data storage and calculation\cite{ref-16}, database management layer, data service layer, data analysis layer, integrated application and terminal layer, and to develop corresponding data standards and application specifications and security Management system, the overall structure as shown (see Fig.\ref{fig-3}).

\subsubsection{Data source layer}
Data sources include the original database, in-line data and external third-party data sources (space-time data, internet data, etc.). In-line data mainly for customer information, transaction information and product information, mainly from the insurance company's core system.

\subsubsection{Data acquisition and convergence layer}

Through the data collection station to collect relevant data, through the data crawler to collect Internet data, remote sensing image analysis and processing technology to obtain the remote sensing geographic information data associated with the insurer, and then through the data aggregation platform to achieve the controllable transmission of information, smooth data channels to achieve Unified collection of data resources, unified management.

\subsubsection{Data storage and calculation layer}
The storage tier includes cluster distributed file system HDFS under Hadoop, distributed memory system Tachyon, and column database HBase \cite{ref-8}. Under the existing safety management system, data and information can be well stored. Computing layer includes processing computing framework MapReduce, DAG computing framework Spark, SQL query framework Hive, machine learning algorithm library Mahout, Mlib and so on\cite{ref-8}. 

\subsubsection{Database management}

Establish the basic database for global application and decision support, including the subject database, basic database and remote sensing spatio-temporal database for insurance application. The data resources are directly related to the business so as to realize the recommendation to potential customers, service tracking, anti-fraud, etc. features.
\subsubsection{data service layer}

The service layer mainly includes the data service support platform to extract data from the big database, and carry out related data processing and interface encapsulation to provide data support services for big data mining and visual analysis to meet various business sectors in the analysis of various types of data The demand.

\subsection{Data Analysis Layer}

Data analysis layer includes space-time big data mining sub-platform and visual analysis sub-platform two parts. Big Data Mining is based on the platform's big data warehouse and provides data mining and analysis capabilities such as data preprocessing, data exploration, model design, model display and model evaluation. It also provides product selection, product recommendation, product pricing, anti-fraud and other areas to provide information support\cite{ref-13}. Visual analysis sub-platform integrated use of graphics analysis, cognitive analysis, correlation analysis and other technologies, the big data to multi-dimensional data representation, to provide sales to different dimensions (including space-time dimension) data observation, and then the data more In-depth observation and analysis.
\subsection{Comprehensive application and terminal layer}

The integrated application and terminal layer includes marketing centers (real-time marketing, social network marketing, event marketing), product centers (product design, product promotion, product pricing, service tracking), operations centers, risk management centers End User Center\cite{ref-9}.

The big data platform is very much used in the financial industry, but for small and medium-sized businesses, the big data platform does not help them to take advantage of it, and large companies have a lot of precious data to lose their chances of competition. Space-time big data platform is currently less used, has a certain potential and forward-looking, if SMEs with their own actual development of their own space-time data platform fit, I believe in a sense, small and medium enterprises can and large Business piece of cake. 

This platform is under design and its architecture (pictured above) has been initially completed, but there is still a lack of research and fieldwork. I hope that in the future of learning, make this platform more complete.

\section{The platform deployment strategy}

The deployment of big data platform is in line with the new model of "Platform + Big data + Integrated Services" and is combined with GIS technology to help break through barriers in traditional industries and become a model in Internet finance, thus forming an exemplary effect and promoting the development of real economy \cite{ref-9}.

Financial Technology focuses more on the application and promotion of information technologies such as big data, cloud computing and GIS, and emphasizes their important role in enhancing financial efficiency and optimizing financial services. The combination of big data platform and GIS technology can reduce costs and prevent risks to a certain extent\cite{ref-11}. More importantly, the formation of a geographically-based customer circle that is not currently existed has fundamentally improved the bank's business capabilities.

\section{Conclusion}

Big data platform combined with space-time data can really integrate its own data resources, but also can be derived from a large number of products, on the cost reduction, risk prevention on the business are helpful. Take the insurance industry as an example. The help of this big data platform for product design, product recommendation, product pricing, product marketing and anti-fraud can be enormous. It can also be said that "financial technology" helps the insurance industry\cite{ref-12}. GIS technology has been quite complete. With the maturing of big data technology, the sparks collide with each other are incalculable\cite{ref-10}.


\begin{acks}
Thanks to School of Entrepreneurship of Heilongjiang University for the support of this project, project number is 201810212005.

\end{acks}
\bibliographystyle{acm}
\bibliography{ref}